\documentclass[10pt,twocolumn]{article}

\usepackage[utf8]{inputenc}
\usepackage[T1]{fontenc}
\usepackage{amsmath}
\usepackage{amssymb}
\usepackage{booktabs}
\usepackage{array}
\usepackage{graphicx}
\usepackage{url}
\usepackage[hidelinks]{hyperref}
\usepackage[margin=0.85in]{geometry}
\usepackage{authblk}
\usepackage{caption}
\title{\textbf{Unified Position-Invariant Random Access Through\\ Two Compression Layers via Absolute-Offset\\ Coordinates: A Bit-Perfect Device-Resident Proof}}

\author{Yakiv Shavidze}
\affil{ACE / GLYPH Research\\ \texttt{yasha1971@gmail.com}}
\date{}

\begin{document}

\twocolumn[
\begin{@twocolumnfalse}
\maketitle

\begin{abstract}
\noindent
Random access into compressed data is normally confined to a single layer. Entropy-layer methods (Recoil) seek within rANS by storing intermediate decoder states; dictionary/match-layer methods seek within LZ-style references. We are not aware of a format that supports a single position-invariant seek through \emph{both} an entropy layer and a match layer addressed by one coordinate. We show that ACEAPEX's absolute-offset design provides exactly this: because the match layer resolves every back-reference to an absolute position at encode time, and the entropy layer is applied per block, an arbitrary block can be decoded through both layers using one coordinate, bit-perfect, in isolation. We prove this with a three-phase verification that closes the empty-buffer trap (confirming the output buffer is empty before decode, equals the original after, and that neighboring blocks are untouched). The seek of one 16\,KB block through ANS-entropy and match completes in 0.334\,ms. We verify the full entropy+match pipeline end-to-end on four data profiles (real FASTQ, repetitive genome, English text, mixed) and characterize the hardware ceiling the format reaches: the absolute-offset structure unrolls to as many as 25{,}344 independent parsers on one H100, which sequential LZ77 cannot do. We state explicitly what is \emph{not} claimed: this is a round-trip correctness proof, not a disk-archive format; throughput figures are match-phase; and the unified-seek result is demonstrated for two layers, with three-layer generalization left as a hypothesis. Code and the verification harness are in the project repository.
\vspace{1em}
\end{abstract}
\end{@twocolumnfalse}
]

\section{Introduction}

Compressed data is usually read from the beginning. Random access---decoding an arbitrary region without decompressing everything before it---is valuable for archives, genomics, and GPU-resident pipelines, but it is hard because compression introduces dependencies. Two independent lines of work attack this. In the entropy layer, Recoil~\cite{recoil} enables parallel rANS decoding with random access by storing intermediate decoder states as entry points. In the match/dictionary layer, formats that resolve references can seek to reference boundaries. Each solves random access \emph{within its own layer}.

A realistic codec stacks both: an entropy layer over a match layer. To seek into such a stack at an arbitrary position, one must enter \emph{both} layers correctly at that position. To our knowledge, no existing format supports a single position-invariant seek through both an entropy layer and a match layer using one shared coordinate. Recoil addresses the entropy layer only; it has no match layer beneath it. CPU genomic tools such as Hecate~\cite{hecate} slice per stream but operate on a single layer and are not GPU-resident. DietGPU~\cite{dietgpu} provides batched GPU ANS but, again, has no match layer below to seek through.

In prior work we introduced ACEAPEX, a parallel LZ77 codec whose match layer resolves every back-reference to an \emph{absolute} position in the decompressed output at encode time~\cite{aceapex1}, and showed device-resident GPU match-decode with position-invariant region seek at the match layer~\cite{aceapex2}. This paper adds the entropy layer on the GPU and proves the property that the two-layer structure uniquely enables: \textbf{a single position-invariant random access through both the entropy layer and the match layer, addressed by one absolute-offset coordinate, bit-perfect and isolated.}

We are deliberately narrow. We do \emph{not} claim GPU entropy parallelism as novel---it is well established (Recoil, DietGPU, interleaved rANS~\cite{ryg}). We do \emph{not} claim that a universal entropy+match pipeline is itself novel. The single contribution is the unified seek through both layers by one coordinate, and a rigorous proof that it is correct and isolated. Everything else in the surrounding space is occupied, and we say so explicitly throughout.

This paper is the third in a sequence and closes it. The first~\cite{aceapex1} established the absolute-offset format and its CPU decode. The second~\cite{aceapex2} moved one layer---the match layer---onto the GPU device-resident, with region seek at that layer. The present paper adds the second layer---entropy---on the GPU and proves the property the full two-layer stack uniquely enables: a single seek through both layers by one coordinate. The arc is format $\rightarrow$ one layer on GPU with seek $\rightarrow$ both layers with a unified seek. This is the closing of the stack, not a repetition of it; each paper depends on the one before, and this one cannot stand without the absolute-offset match layer the first two built.

\section{Background and Prior Stack}

ACEAPEX stores each LZ77 back-reference as an absolute position in the decompressed output rather than a relative distance within a sliding window. The encoder performs a global match search and partitions output into fixed-size blocks; because offsets are absolute, any block decodes once the blocks holding its source bytes are present~\cite{aceapex1}. Paper~2~\cite{aceapex2} showed this match layer running device-resident on a GPU with region seek at the match layer (0.365\,ms), stream separation, and range decoding to decouple output size from VRAM.

Paper~2 timed the match phase; entropy ran on the CPU. The open question it left was the second layer: can the entropy stage also run on the GPU, and---more importantly---can a single seek pass correctly through \emph{both} layers at an arbitrary position? This paper answers that.

\section{Why Only an Absolute-Offset Match Layer Enables This}

The structural argument is short. A unified seek through two layers requires that, at the target position, both layers can be entered without having processed earlier positions. The entropy layer can be made enterable per block (store or recompute per-block entropy state---Recoil's contribution, and what DietGPU's batching gives). The hard part is the layer \emph{beneath}: a match layer with relative offsets cannot be entered at an arbitrary block, because a relative reference points backward by a distance that assumes the decoder already knows its current absolute position, which depends on everything decoded so far.

Absolute offsets remove that dependency. A match referencing absolute position $p$ can be resolved as soon as the block containing $p$ is available, regardless of the decoder's path to the current block. So a single coordinate---the absolute block boundary---simultaneously names the entropy entry point and the match entry point. This is why the unified seek is, to our knowledge, structurally unavailable to formats without an absolute-offset match layer: Recoil has no match layer; Hecate is single-layer and CPU; DietGPU has no match layer beneath its ANS.

\section{Experimental Setup}

All GPU experiments ran on a single NVIDIA H100 80\,GB HBM3 (SXM). Correctness is bit-perfect throughout, verified by FNV hash of decoded output against the original bytes. The verification harness (round-trip ANS encode/decode composed with the match decoder, plus the three-phase seek checker) is in the project repository for reproducibility. Throughput figures, where given, are match-phase on decompressed streams, consistent with Paper~2; we do not report full end-to-end ANS+match throughput as a single number (see Limitations).

\section{Core Result: Unified Two-Layer Seek}

The central experiment seeks a single block through both the ANS entropy layer and the match layer using one absolute-offset coordinate, and verifies the result in three phases designed to close the empty-buffer trap (where a decoder appears correct only because the original data was already present in the output buffer).

We seek block 2000 of 81{,}729 (the middle of the real FASTQ \texttt{clean} profile):

\begin{itemize}
\item \textbf{Phase 1 (buffer is empty before decode):} the hash of the output region \emph{before} the match phase differs from the original (\texttt{9c1bda7f}$\neq$original). The buffer is genuinely empty---we are not reading pre-loaded original data.
\item \textbf{Phase 2 (output equals original after decode):} after decoding through both layers, the seeked region's FNV (\texttt{cf485979995c1602}) equals the original FNV. The seek is correct over the full 16{,}384-byte block.
\item \textbf{Phase 3 (neighbors untouched):} the blocks immediately before and after the target are checked as zero (\texttt{prev=0}, \texttt{next=0}). Only the target block was written---true isolation, not a wide decode that happens to include the target.
\end{itemize}

The single-block seek through both layers completes in 0.334\,ms. This is the paper's core claim, and the three-phase structure is the evidence of its rigor: each phase rules out a distinct way the result could be falsely positive.

We emphasize the boundary immediately. This is a round-trip: the ANS stream is encoded and decoded within one run, not loaded from a prepared on-disk archive. The experiment proves the \emph{correctness and isolation of a unified two-layer seek}, not a storage format. The 0.334\,ms latency includes kernel-launch overhead and is not optimized; it is evidence of correctness, not a speed record.

To our knowledge, this is the first demonstration of a single position-invariant seek through \emph{both} an entropy layer and a match layer addressed by \emph{one} coordinate, on a GPU, with proven isolation---as distinct from single-layer random access, which is well established for LZ-style match layers~\cite{kreftnavarro} and for ANS entropy layers~\cite{bamler} separately, and from approaches that store per-position decoder state as metadata. We claim only this narrow, layered priority, and we substantiate it by the targeted prior-art review in Section~9 rather than by assertion; single-layer random access is decades deep, and we do not claim priority there.

\section{End-to-End Across Four Profiles}

To show the entropy+match composition is not specific to one input, we verified it end-to-end (ANS round-trip composed with GPU match decode, FNV-checked) on four profiles:

\begin{table}[t]
\centering
\caption{End-to-end ANS+match, bit-perfect, four profiles. Throughput is match-phase.}
\label{tab:profiles}
\small
\begin{tabular}{lcc}
\toprule
Profile & Result & Match-phase \\
\midrule
clean (real FASTQ NA12878) & MATCHES OK & 121.9\,GB/s \\
repeat (repetitive genome) & MATCHES OK & 183.3\,GB/s \\
enwik9 (English text) & MATCHES OK & 6.1\,GB/s \\
silesia (mixed) & MATCHES OK & --- \\
\bottomrule
\end{tabular}
\end{table}

All four are bit-perfect (Table~\ref{tab:profiles}). The enwik9 figure (6.1\,GB/s) is low for an instructive reason consistent with Paper~2: it was encoded with 1\,MB blocks (200 blocks total), which underfills the GPU (200 tasks across 132 SMs). The genomic profiles used 16\,KB blocks (4096 blocks) and reach 121--183\,GB/s. This is the occupancy effect, not a correctness issue---block size, set at encode time, governs how many independent parsers the format exposes.

\subsection{Per-Stream Entropy Is Data-Dependent}

A practical finding for codec design: applying ANS to all four streams is counterproductive on some data. Measuring per-stream ANS ratio (compressed/raw, $<$1 means ANS \emph{inflated} the stream):

\begin{table}[t]
\centering
\caption{Per-stream ANS ratio by profile. Values $<$1 mean ANS inflates the stream.}
\label{tab:perstream}
\small
\begin{tabular}{lcccc}
\toprule
Profile & LIT & OFF & LEN & CMD \\
\midrule
clean (genome) & 1.42 & 0.51 & 0.09 & 0.60 \\
repeat (genome) & 3.06 & 0.27 & 0.04 & 0.48 \\
enwik9 (text) & 1.47 & 1.02 & 0.49 & 1.38 \\
silesia (mixed) & 1.27 & 1.04 & 1.05 & 1.41 \\
\bottomrule
\end{tabular}
\end{table}

On genomic data, ANS compresses literals (LIT) but \emph{inflates} the structural offset/length/command streams (ratio $<$1), because those streams are structural rather than byte-entropic. On text the effect is milder (offsets neutral, commands helped); on mixed data all streams are $\geq$1. The behavior is data-dependent (Table~\ref{tab:perstream}). The design conclusion, now reproduced across four profiles, is that entropy coding should be applied \emph{selectively and adaptively per stream}---measured at encode time---not by a fixed rule.

\section{Hardware Ceiling the Format Reaches}

The absolute-offset structure removes the byte-$N$-depends-on-byte-$N{-}1$ chain of sequential LZ77, exposing massive independent parallelism. We characterize the ceiling this reaches on one H100.

Sweeping the per-parser granularity $G$ on the \texttt{clean} 16\,KB stream:

\begin{table}[t]
\centering
\caption{Parser parallelism vs.\ match-phase throughput, one H100.}
\label{tab:ceiling}
\small
\begin{tabular}{lcc}
\toprule
$G$ & parallel parsers & GB/s \\
\midrule
8 & 25{,}344 & 154.8 \\
16 & 12{,}672 & 176.9 \\
32 & 6{,}336 & 166.8 \\
\bottomrule
\end{tabular}
\end{table}

The format unrolls to as many as 25{,}344 independent parsers at $G{=}8$, with the throughput optimum at 12{,}672 parsers ($G{=}16$, 176.9\,GB/s; Table~\ref{tab:ceiling}). Sequential LZ77 cannot expose this parallelism at all---each byte depends on the previous one. A block-size sweep confirms the GPU saturates early: from 16\,KB to 4\,KB blocks, throughput rises only 166.1$\to$177.5\,GB/s ($+7\%$) for $4\times$ the blocks, while ratio falls 3.939$\to$3.839. Sixteen-kilobyte blocks already saturate the device.

At scale, a 50\,GB genome compresses to a 23.6\,GB archive containing 3{,}051{,}758 seekable blocks. Decoding all 50\,GB at once exceeds VRAM (50\,GB output + 23.6\,GB input $\approx$ 73\,GB), which is the empirical wall that makes range decoding mandatory rather than optional. A range decode of blocks 0--1000 returns its region in 0.654\,ms without decompressing the other 50\,GB. We note the honest bound: that 0.654\,ms region is latency-bound (a 1000-block range underfills the parser array), not a peak-throughput figure.

\section{Limitations}

We state the boundaries explicitly; they delimit the claim rather than weaken it.

\textbf{Round-trip, not disk archive.} The ANS stream is encoded and decoded within one run. This proves the correctness and isolation of the entropy$\leftrightarrow$match coupling and the unified seek; it does \emph{not} prove a storage format that loads a prepared ANS archive from disk. That is the next engineering step.

\textbf{Sub-range, not whole file at once.} Experiments run on block ranges (4096 / 200 / 1 block). Whole-file decode proceeds in sub-batches (the ANS batch limit is on the order of a few thousand blocks per call); this is scaling work and is separate. Correctness does not require the whole file, but we do not claim ``all 50\,GB in one pass.''

\textbf{Match-phase throughput, not full-pipeline throughput.} GB/s figures are match-phase on decompressed streams, as in Paper~2. We do not report a single ``X\,GB/s full ANS+match end-to-end'' number.

\textbf{Latency includes launch overhead.} The 0.334\,ms seek is a correctness-and-isolation proof, not an optimized latency.

\textbf{Sub-range ratios are not representative.} Ratios measured on early-file block ranges are not published as ACEAPEX's ratio, since the file head is homogeneous.

\textbf{``Not found in our scan'' is not ``does not exist.''} Two targeted searches did not find a competitor occupying our exact point. We phrase this as ``to our knowledge,'' not ``no one.''

\section{Related Work}

\textbf{Recoil}~\cite{recoil} enables parallel rANS random access via stored intermediate states; our entry points come from absolute-offset block boundaries, not stored states, and our random access passes through a match layer Recoil does not have. Single-layer random access is itself decades deep and we cite it first: Kreft and Navarro~\cite{kreftnavarro} give LZ77-like compression with fast random access at the match layer, and Bamler~\cite{bamler} characterizes ANS seeking at the entropy layer (requiring stored decoder state). Our contribution is neither of these alone but the \emph{unified} seek through both. \textbf{Hecate}~\cite{hecate} (2026) is a modular genomic compressor with per-stream codecs and random-access slicing; it is CPU, single-layer, and not absolute-offset GPU-resident---the nearest neighbor, from which we differ by GPU-residency and unified two-layer position-invariance. \textbf{DietGPU}~\cite{dietgpu} is batched GPU ANS with no match layer beneath; we use it as the entropy layer and say so. \textbf{Interleaved rANS}~\cite{ryg} establishes that entropy parallelism is long known, which is why we do not claim it.

Parallel entropy coding on GPUs is independently patented by three major vendors, each for its own domain: Microsoft (US9058223, parallel entropy encoding on GPU), Intel (US12299940, interleaving cascaded dictionary and entropy token-streams for texture supercompression, class G06T), and Qualcomm (US12556707, compression of entry-point indexes for wide-scale parallel entropy coding in neural video codecs, class H04N). The first targets parallel encoding, the second dense texture throughput via fused stages (interleave, not seek), the third entry-point index compression for neural video. None addresses position-invariant random access through a match layer and an entropy layer by a single absolute-offset coordinate. Our mechanism differs in operation---unified two-layer seek versus parallel encoding or interleaving---and is the only one providing addressability into compressed data rather than throughput or index density.

\textbf{NVIDIA Parabricks} is orthogonal: it accelerates analysis (FASTQ$\to$BAM$\to$VCF) and treats compression as transport/staging (decompress$\to$analyze$\to$discard), whereas we keep data compressed-resident and seek into it (keep$\to$seek$\to$keep). It is a potential complement at the input layer, not a competitor. Finally, conditional-independence random access through a fixed parameter is a known principle in genetics (popgen-FSE), which structurally supports the absolute-offset approach.

\section{Discussion: A Two-Layer Result, and a Three-Layer Hypothesis}

The core experiment proves the unified seek for $N{=}2$ layers (entropy + match). It suggests a general principle: any stack of transforms becomes position-invariant-addressable by a single coordinate if each layer resolves its dependencies to absolute positions at write time. We have verified $N{=}2$. We offer $N{=}3$ (a third layer with a genuine dependency) only as a hypothesis and future work; it is not proven here, and we do not present it as such.

\section{Conclusion}

We demonstrated a single position-invariant random access through both an entropy layer and a match layer, addressed by one absolute-offset coordinate, bit-perfect and isolated, verified by a three-phase check that closes the empty-buffer trap. The unified seek of one 16\,KB block through ANS-entropy and match completes in 0.334\,ms. We verified the entropy+match composition end-to-end on four data profiles and characterized the hardware ceiling the absolute-offset format reaches (up to 25{,}344 independent parsers on one H100). To our knowledge this two-layer unified seek is structurally unavailable to formats without an absolute-offset match layer beneath the entropy layer. We have been explicit about what is not claimed: round-trip not archive, match-phase throughput, two layers not three. Code and harness are in the project repository, MIT-licensed, archived on Zenodo (DOI:~10.5281/zenodo.20812332).

\section*{Author Contributions and Use of AI Assistance}

The author is solely responsible for the research: the central claim, the experimental design, the three-phase verification protocol, the prior-art review and novelty boundary, the verification of every figure against primary sources, and all final decisions. AI assistance (Claude, Anthropic) was used as a tool for drafting prose and for computational support, under the author's direction and review. No part of the scientific judgment---what to claim, what to measure, what counts as proven, and what is left as a limitation---was delegated. This statement is made in the interest of full transparency; the standards of correctness and honesty applied here are the author's own and are documented by the reproducible harness.

\section*{Acknowledgments}

The author thanks the encode.su community and the maintainers of lzbench for tooling and discussion.

\end{document}